\documentclass{article}

\usepackage[preprint,nonatbib]{neurips_2019}

\usepackage[utf8]{inputenc} 
\usepackage[T1]{fontenc}    
\usepackage{hyperref}       
\usepackage{url}            
\usepackage{booktabs}       
\usepackage{amsfonts}       
\usepackage{nicefrac}       
\usepackage{microtype}      
\usepackage{blindtext}
\usepackage{amsmath,amssymb}
\usepackage{graphicx}
\usepackage{algorithm}       
\usepackage[noend]{algpseudocode} 
\usepackage{multirow}
\usepackage{dcolumn}
\usepackage{subcaption}

\title{A Dual-Hormone Closed-Loop Delivery System for Type 1 Diabetes Using Deep Reinforcement Learning 
}

\author{%
   Taiyu Zhu \\
   Imperial College London \\
   London, SW7 2AZ \\
   \texttt{taiyu.zhu17@imperial.ac.uk} \\
   \And
     Kezhi Li\thanks{ T. Zhu and K. Li have equal contribution} \\
  Imperial College London\\
  London, SW7 2AZ \\
  \texttt{kezhi.li@imperial.ac.uk} \\
   \AND
   Pantelis Georgiou \\
   Imperial College London \\
    London, SW7 2AZ \\
   \texttt{pantelis@imperial.ac.uk} \\
}

\begin{document}

\maketitle

\begin{abstract}


We propose a dual-hormone delivery strategy by exploiting deep reinforcement learning (RL) for people with Type 1 Diabetes (T1D). Specifically, double dilated recurrent neural networks (RNN) are used to learn the hormone delivery strategy, trained by a variant of Q-learning, whose inputs are raw data of glucose \& meal carbohydrate and outputs are dual-hormone (insulin and glucagon) delivery. Without prior knowledge of the glucose-insulin metabolism, we run the method on the UVA/Padova simulator. Hundreds days of self-play are performed to obtain a generalized model, then importance sampling is adopted to customize the model for personal use.
\emph{In-silico} the proposed strategy achieves glucose time in target range (TIR) $93\%$ for adults and $83\%$ for adolescents given standard bolus, outperforming previous approaches significantly. The results indicate that deep RL is effective in building personalized hormone delivery strategy for people with T1D.




\end{abstract}

\section{Introduction}

Diabetes is a lifelong condition that affects an estimated 425 million people worldwide \cite{IDF-2017}.
Delivering an optimal insulin dose to T1D subjects is one of the long-standing challenges since 1970s \cite{Lunze-ArtificialPan2013, Reddy-CliSafety2016}.
Previously the quality of life of T1D subjects heavily relies on the accuracy of human-defined models \& features of the delivery strategy.
In recent years, deep learning (DL) has provided new ideas and solutions to many healthcare problems \cite{jiang-artificial2017}, empowered by increasing availability of medical data and rapid progress of analytic tools, e.g. RL.
However, several reasons hinder us from building an efficient RL model to solve problems in chronic diseases.
Firstly, collected from a dynamic interaction between human and environment, RL medical data are limited and expensive \cite{Artman-power2018}.  In addition, different from playing Atari in virtual environment \cite{Mnih-playingAtari2013}, RL costs heavily to 'explore' possibilities on human beings in terms of prices and safeness.
Finally, the variability of physiological responses to the same treatment can be very large for different people with T1D \cite{Vettoretti-Type1Dia2018}. It is difficult to build a versatile model suitable for wide subjects. 
These reasons lead to little progress of using RL in chronic illnesses.

To overcome the obstacles, we propose a two-step framework to apply deep RL in chronic illnesses, and use T1D as a study case. T1D is chosen because it is a typical disease that requires dynamic treatment consistently. A generalized DNN for hormone delivery strategy is pre-trained using a variant of Q-learning as the first step. We use double dilated recurrent neural network (DRNN) to build a model for multi-dimensional medical time series, in which each basal hormone delivery (at five minutes intervals) are considered as an action determined by a stochastic policy, and glucose levels and TIR are considered as the reward.
Secondly, by adopting weights \& features from the last step, important sampling are implemented to customize the model personalized using individual data.


We use the UVA/Padova T1D Simulator, a popular glucose-insulin dynamics simulator which has been accepted by the Food and Drug Administration (FDA) \cite{Man-UVA/PADOVA2014}, as the environment. It can generate T1D subjects data with high variability in meals, body conditions and other factors.
During the training, hundreds days of 'self-play' trials are performed to obtain a generalized model for the dual-hormone closed-loop basal delivery with standard bolus.
In the test, 10 adults and 10 adolescents are tested within 6 months period of time. The results show that TIRs achieve $93\%$ for adults and $84\%$ for adolescents \emph{in-silico}, which significantly improve the state-of-the-art performances. 



\section{Related Work and Preliminaries}
The rapid growth of continuous glucose monitoring (CGM) and insulin pump therapy have motivated use of a closed-loop system, known as the artificial pancreas (AP) \cite{cobelli-artificial2011,Hovorka-CloseLoop2011,russell-outpatient2014,herrero-coordinated2017}. Many algorithms are developed and verified as closed-loop single/dual hormone delivery strategies  \cite{bergenstal-SafetyOf2016} that are mostly based on control algorithms \cite{Facchinetti-ConGlu2016,Haidar-TheAP2016}.  Machine learning (ML) is leveraged in glucose management in \cite{Plis-AMachine2014,Mhaskar-ADeepLearningApp2017}. DNN is a new method in glucose management with fully connected \cite{perez-ArtiNN2010}, CNN-based \cite{Li-CRNN2019}, RNN-based \cite{Chen-DilatedRec2018}, physiological-based \cite{Bertachi-PreofBlo2018} networks. It has been shown that dilated RNN performs well in processing long-term dependencies \cite{chang-dilated2017}. By updating control parameters in gradient, RL is adopted recently \cite{Ngo-ControlOfBl2018,herrero-coordinated2017,Sun-ADual2019}.
To accelerate the learning process, prioritized experience replay samples important transitions more frequently \cite{schaul-prioritized2015}.
Before submission we find an RL environment was built in simulator of 2008 version \cite{Jinyu-Simglucose2018}, while we use a simulator of 2012 version.

We see the problem an infinite-state Markov decision process (MDP) with noises. An MDP can be defined by a tuple $\langle S, A, R, \mathcal{T},  \gamma \rangle$ with state $S$, action $A$, reward function $R: S\times A  \mapsto [R_{\text{min}},R_{\text{max}}]$, transition function $\mathcal{T}$, and discount factor $\gamma  \mapsto [0,1]$. At each time period, the agent takes an action $a\in A$, causes the environment from some state $s\in S$ to transit to state $s'$ with probability $T(s', s,a) = P(s'|s,a)$. A policy $\pi$ specifies the strategy of selecting an action. RL's goal is to find the policy $\pi$ mapping states to actions that maximizes the expected long-term reward. A Q-function $Q^{\pi}(s,a)$ can be defined to computed this reward $Q^{\pi}(s,a) = R(s,a) + \gamma \sum_{s'}P(s'|s,a)\sum_{a'}\pi(s',a')Q^{\pi}(s',a')$. The optimal action-value function $Q^*(s,a)=\max_{\pi}Q^{\pi}(s,a)$ offers the maximal values that can be determined by solving the Bellman equation $Q^*(s,a) = \mathbb{E}\left[ R(s,a) + \gamma\sum_{s'}P(s'|s,a) \max_{a'}Q^*(s',a')\right]$.


\section{Methods}


In the hormone delivery problem, we use a multi-dimensional data as the input $D$. The data includes blood glucose ${G}$ (mg/dL), meal data ${M}$ (g) manually record by individuals, corresponding meal bolus ${B}$ (U).  Dual-hormone (basal and glucagon) delivery is considered as actions ${A}$.  In this case $D$ can be denoted as $D =\{ G,M,I,C\} = [d_1, \cdots,d_L]^\mathbb{T} \in \mathbb{R}^{\mathbb{L}\times 4}$ where $L$ is the data length, $I$ is the total insulin including bolus $B$ and basal plus the glucagon $C$. We use the latest 1 hour data (12 samples) as current state $s_t = [d_{t-11}\cdots,d_{t-1},d_t]^\mathbb{T}$. 
Here $B$ is computed from $M$ with a standard bolus calculator $B \propto M$ divided by the body weight.
Then the problem can be seen as an agent interacting with an environment over time steps sequentially. Every five minutes, an observation $o_t= s_t+e_t$ can be obtained, and an action $a_t$ is taken. The action can be chosen from three options: do nothing, deliver basal insulin, or deliver glucagon. The amount of basal insulin and glucagon is a small constant that determined by the subject profile in advance. To maintain the BG in a target range, we define a reward carefully that the agent receives in each time step
\begin{equation}
r_t =     \left\{\begin{array}{l}1 \ \ \ \qquad \qquad\qquad\qquad\qquad  90\leq G_t\leq 140\\ 0 \ \ \ \qquad\qquad\qquad\qquad\qquad 70\leq G_t< 90 \ \& \  140< G_t\leq 180 \\ -0.4-(G_t-180)/200 \ \ \ \ \ 180< G_t\leq 300 \\ -0.6+(G_t-70)/100 \ \ \ \ \ \ \ 30\leq G_t< 70 \\ -1  \qquad\qquad\qquad\qquad\qquad \text{else} \end{array}\right.
\label{eq:reward}
\end{equation}
The goal of the agent is to learn a personalized basal insulin and glucagon delivery strategy in a short period of time (using limited data) for each individual. Here we propose a two-step learning approach to build the dual-hormone delivery model.


\subsection{Generalized DQN training}


We build an interactive environment in \emph{simulator}, and dilated RNN is employed as the DNN. Dilated RNN is used because it has larger receptive field that is crucial for glucose time series processing \cite{Chen-DilatedRec2018}. Double DQN weights $\theta_1,\theta_2$ in the \emph{simulator} are trained
because it has been proved as a robust approach to solve overestimations. Action network and value network are trained $J_{DQ}(Q) = \left( R(o,a) + \gamma Q(o', {\operatorname {arg~max} }_a Q(o',a;\theta_1);\theta_2)-Q_i(o,a;\theta_1) \right)^2$.  The pseudo-code is sketched in Algorithm 1.

\begin{algorithm}
\caption{Generalized model training}
\begin{algorithmic}[1]

\State Inputs: Initializing environment $E$ , historical data $H$, update frequency T, two dilated RNN of random weights $\theta_1$, $\theta_2$, respectively.
\Repeat
\State sample action from $a\sim \pi(Q_{\theta_1},\varepsilon)$, observe $(o',r)$ in $E(Is)$
\State store $(o,a,r,o')$ into replay buffer $\mathcal{B}$
\State sample a mini-bath uniformly from $\mathcal{B}$ and calculate loss $J_{DQ}(Q)$
\State perform a gradient descent to update $\theta_1$, $\theta_2$
\State \textbf{if} $t$ \text{mod} $T = 0$ \textbf{then} $\theta_2\leftarrow \theta_1$ \textbf{end if}
\Until{converge or reach the number of iterations}
\end{algorithmic}
\end{algorithm}

The agent explores random hormone delivery actions under policy $\pi$ that is $\varepsilon$-greedy with respect to $Q_{\theta_1}$ in \emph{simulator}. It is similar to playing Atari games, so the agent can 'self-play' in \emph{simulator} for long time.
Some human intervention/demonstration at the beginning of the RL process can reduce the training time slightly, but in our case it is not necessary.
At the end of this step, a DRNN with weights $\theta$ is obtained as the generalized model.

\subsection{Personalized DQN training}

In this step we refine the model and customize it for the personal use. Weights and features obtained from the last step are updated using limited data with an importance sampling \cite{schaul-prioritized2015} and safety constraints.
It gives better actions (leading to no hyperglycemia or hypoglycemia) higher probability and worse demonstrations lower probability.
Details are shown in Algorithm 2.

\begin{algorithm}
\caption{Personalized DQN training using imprtance sampling}
\begin{algorithmic}[1]

\State Inputs: Initialized with environment $E$, historical data $H$, generalized Q-function $Q$ with weights $\theta_1$, replay buffer $\mathcal{B}$, target weights $\theta_2$,  update frequency $T$ 
\State  generate $\mathcal{D}$ as a merge of $\mathcal{B}$ and experience collected from $H$
\State  calculate importance probability $Pr$ from $H$
\Repeat
\State sample action from policy $a\sim \pi(Q_{\theta},\varepsilon)$, observe $(o',r)$
\State store $(o,a,r,o')$ in $\mathcal{D}$, overwriting the oldest samples previously merged from $\mathcal{B}$
\State sample a mini-batch from $\mathcal{D}$ with modified importance sampling $Pr$
\State calculate loss $J(Q) = J_{DQ}(Q)$
\State perform a gradient descent update $\theta$ and the importance
\State \textbf{if} $t$ \text{mod} $T = 0$ \textbf{then} $\theta_2\leftarrow \theta_1$ \textbf{end if}
\Until{converge or reach the number of iterations}
\end{algorithmic}
\end{algorithm}

\section{Experiment Results}
We compare the results with the following experiment setup (details in supplementary materials): 1. bolus calculator + constant basal insulin (BC); 2. bolus calculator + insulin suspension + carbohydrate recommendation (BC+IS+CR) \cite{liu-coordinating2019}; 3. ours:  bolus calculator + generalized RL model (BC+RL) (Algorithm 1); 4. ours:  bolus calculator + RL intelligent basal (BC+IB) (Algorithm 1, 2)

In experiments we used the TIR ($[70,180]$ mg/dL), the percentage of hypoglycemia ($< 70$ mg/dL) and hyperglycemia ($>180$ mg/dL) as the metrics to measure the performance.
In general, either higher TIR or lower Hypo/Hyper indicates better glycemia control. These evaluation metrics are preferred in diabetes clinic  \cite{vigersky2018relationship}.
Table \ref{table_1} presents the overall performance of the experiment setup on the adult and adolescent subjects. For adult subjects, the IB model achieves the best performance, and increases the mean TIR by 11.21\% ($p\leq0.005$) compared with the BC setup.
In the adolescent case, the IB model obtains the best TIR of 83.39\%.
In both cases, the IB model outperforms the controller with IS and CR on both TIR and Hypo/Hyper results with considerable improvements.



\begin{table}[t]
\caption{Results of different experiment setup}
\label{table_1}
\centering

\begin{tabular}{|c|c|c|c|c|}
\hline
(\%)                    & Method   & TIR & Hypo & Hyper \\ \hline
\multirow{4}{*}{Adult}      & BC         &$81.91\!\pm \!8.66^\ddagger$ & $5.29\!\pm \!3.93^\ddagger$ & $12.80\!\pm \!8.67^\ddagger$       \\ \cline{2-5}
                            & BC+IS+CR   & $87.62 \!\pm\! 7.57^\ddagger$ &$2.36\!\pm \!1.44^\ddagger$   & $10.01\!\pm \!7.35^\ddagger$\\ \cline{2-5}
                            & BC+RL      & $89.16\!\pm\!5.04$&$1.92\!\pm\!1.36$ & $8.92\!\pm\!5.38$     \\ \cline{2-5}
                            & BC+IB     & $\textbf{93.12}\!\pm \!\textbf{4.48}$ &  $\textbf{1.25}\!\pm \!\textbf{1.32}$ & $\textbf{5.63}\!\pm \!\textbf{3.29}$   \\ \hline
\multirow{4}{*}{Adolescent} & BC        & $61.68\!\pm\!10.95^\ddagger$ &$9.04\!\pm\!7.22^\ddagger$  & $29.28\!\pm\!11.16^\ddagger$ \\ \cline{2-5}
                            & BC+IS+CR     & $74.55\!\pm\!9.61^\dag$  & $2.38\!\pm\!1.82^\dag$  & $23.07\!\pm\!7.26^\ddagger$      \\ \cline{2-5}
                            & BC+RL   & $74.89\!\pm\!{8.58}$  & $2.36\!\pm\!{2.19}$  & $22.75\!\pm\!8.63$      \\ \cline{2-5}
                            & BC+IB   & $ \textbf{83.39}\!\pm\! \textbf{8.03}$     & $ \textbf{2.10}\!\pm\! \textbf{1.56}$ &  $ \textbf{14.51}\!\pm\! \textbf{9.98}$      \\ \hline
\end{tabular}
 ~

\footnotesize{$^\ast p\leq 0.05$  $^\dag p\leq0.01$  $^\ddagger p\leq0.005$}\\
\end{table}

\begin{figure*}[!t]
    \centering

    \begin{subfigure}[b]{0.5\textwidth}
        \centering
         \includegraphics[height=3.5in]{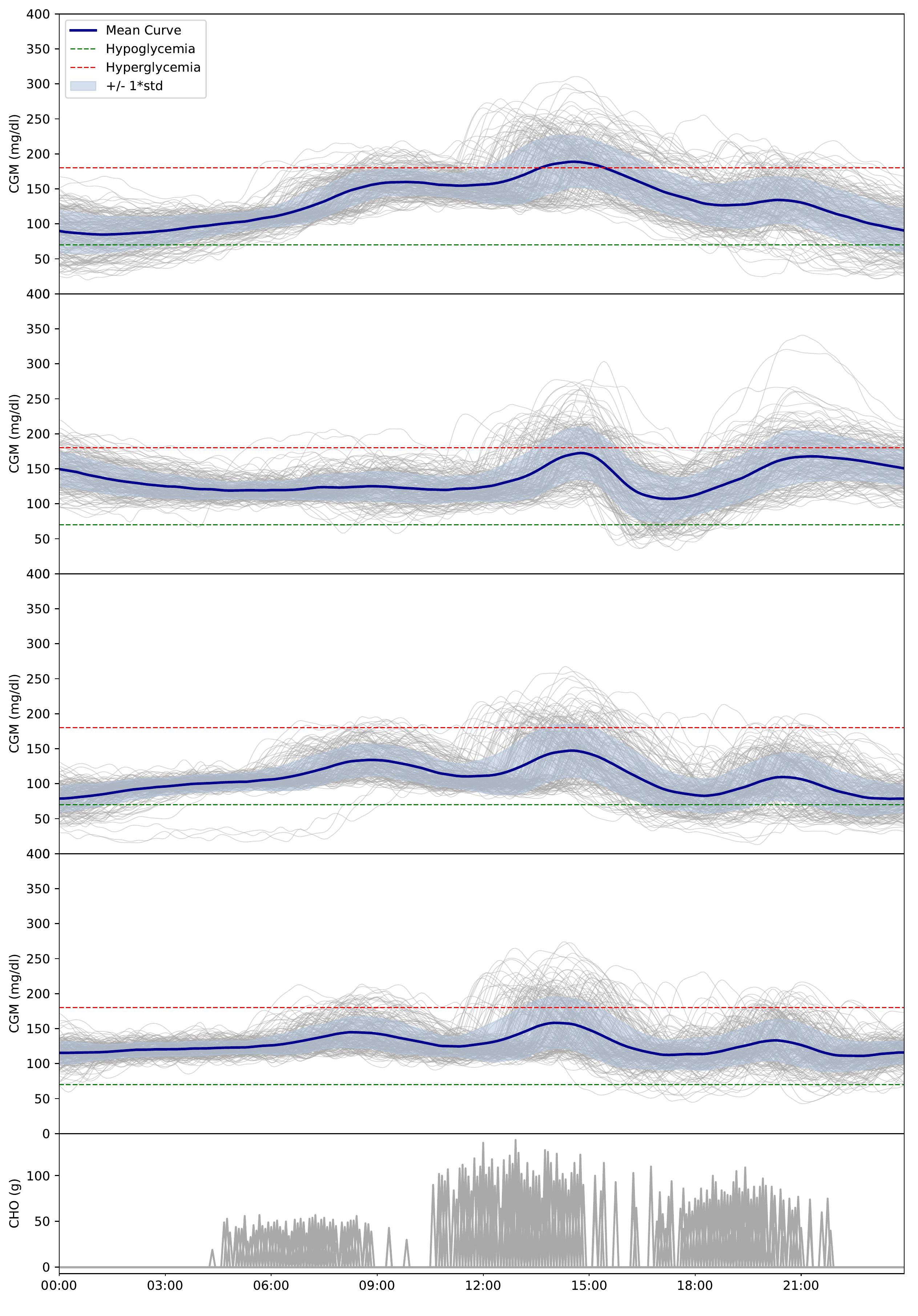}

         \caption{An adult simulation}
    \end{subfigure}%
    ~
    \begin{subfigure}[b]{0.5\textwidth}
        \centering
         \includegraphics[height=3.5in]{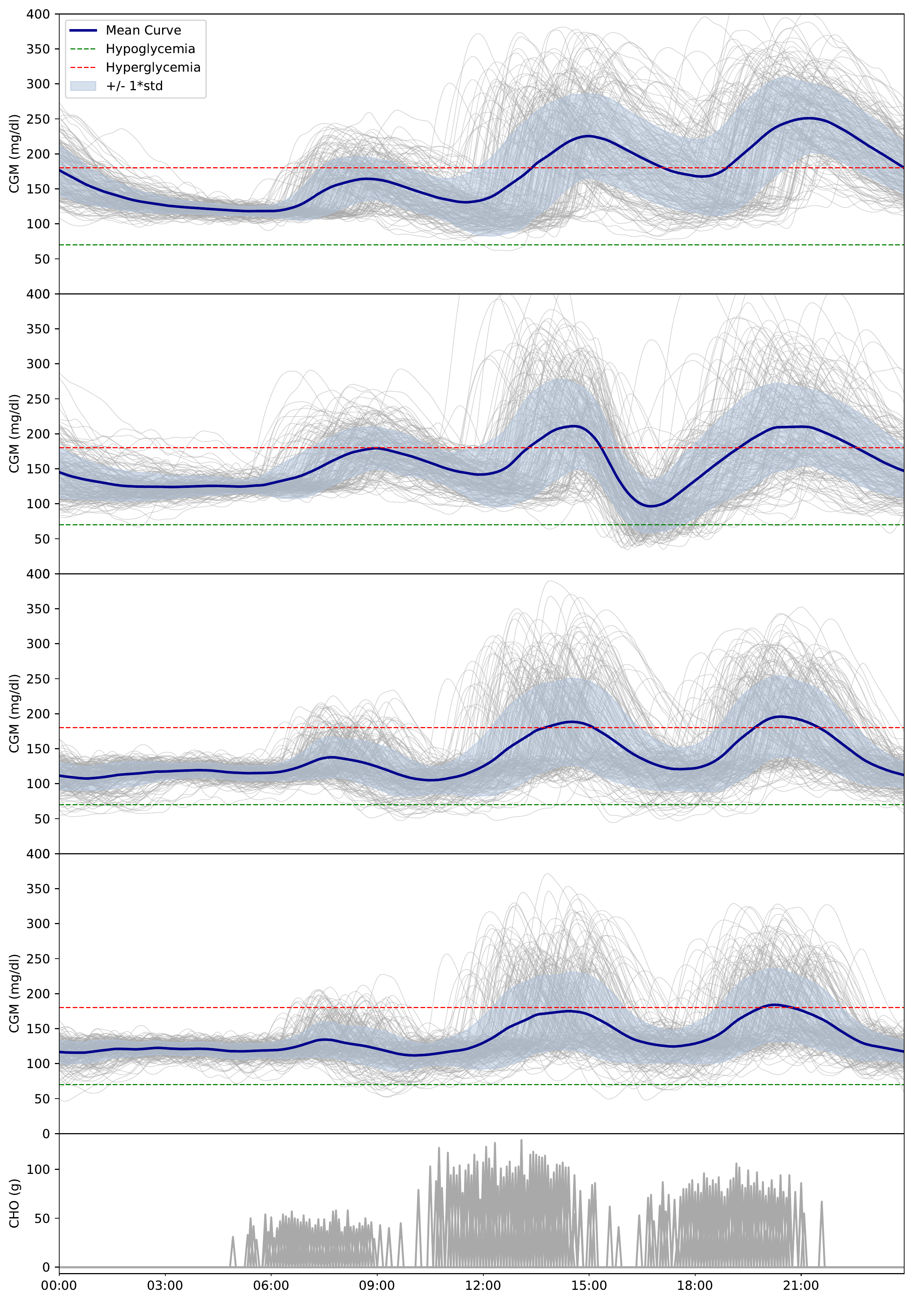}

        \caption{An adolescent simulation}
    \end{subfigure}%

\label{fig:BG_curves}
\caption{Performance of the four experiment setup on an adult subject and an adolescent subject in a 6-month period: (Top-to-bottom) BC, BC+IS+CR, BC+RL (ours), BC+IB (ours), distribution of carbohydrate intake. The average BG levels for 180 days are shown in solid blue lines, and the hypo/hyperglycemia regions are shown in dotted green/red lines. The gray lines correspond to glucose levels in every day of the trial, and the blue shaded regions indicate the standard deviation.}
\end{figure*}

To observe the model performance on the specific BG levels, we visualize the BG curves of two subjects over 6-month testing period in Figure 2.
For the adult curves at the left column, the performance is basically in accordance with statistical results in Table~\ref{table_1}, and
the IB model avoids many hypoglycemia events during the night.
At the right column,
it can be observed that implementation of the IB model significantly helps to reduce the overall BG levels and increase TIR by making the mean curve flat and stable. Surprisingly, standard CNN or LSTM cannot achieve such good (or close) performances as dilated RNN has achieved.

\section{Conclusion}

We propose an intelligent basal hormone delivery algorithm and employ deep RL in glucose management. A dilated RNN is used in double DQN, and a personalized model is obtained from generalized pre-trained model. The results outperform many existing work significantly.
\medskip

\small
\bibliographystyle{unsrt}

\end{document}